\documentclass[twocolumn]{jpsj3}

\usepackage{color}
\usepackage{times}
\usepackage{graphicx}

\title{Two-channel Kondo Effect Emerging from Nd Ions}

\author{Takashi Hotta}

\inst{Department of Physics, Tokyo Metropolitan University,
Hachioji, Tokyo 192-0397, Japan}

\recdate{\today}

\abst{
We discuss Kondo phenomena in a seven-orbital impurity Anderson model
hybridized with $\Gamma_8$ conduction electrons
by employing a numerical renormalization group method.
In particular, we focus on the case with three local $f$ electrons,
corresponding to a Nd$^{3+}$ ion.
For realistic values of Coulomb interactions,
spin-orbit coupling, cubic crystalline electric field potentials,
and hybridization, we find a residual entropy of $0.5\log 2$, a
characteristic of two-channel Kondo phenomena,
for the wide range of parameters of the local $\Gamma_6$ ground state.
This is considered to be the magnetic two-channel Kondo effect,
consistent with the result from an extended $s$-$d$ model
constructed on the basis of the $j$-$j$ coupling scheme.
Finally, we briefly discuss candidates of Nd compounds
to observe the two-channel Kondo effect.
}

\begin{document}
\maketitle

It is one of the fascinating problems in the modern condensed matter physics
to realize an exotic new quantum state in strongly correlated electron systems.
Among them, concerning the non-Fermi liquid state, the
two-channel Kondo effect has been discussed for a long time
as it is a confirmed route to arrive at the non-Fermi liquid ground state.
Coqblin and Schrieffer derived exchange interactions
from the multiorbital Anderson model.\cite{Coqblin}
Then, the concept of the multichannel Kondo effect was developed
on the basis of such exchange interactions,\cite{Nozieres}
as a potential source of non-Fermi liquid phenomena.
Moreover, such non-Fermi liquid properties were pointed out
in a two-impurity Kondo system.\cite{Jones1,Jones2}

Concerning the reality of two-channel Kondo phenomena,
Cox pointed out the existence of two screening channels
in the case of quadrupole degrees of freedom in a cubic U compound
with a $\Gamma_3$ non-Kramers doublet ground state.\cite{Cox1,Cox2}
As Cox's idea attracted significant attention,
a large number of works were published on this topic
and the understanding on the two-channel Kondo phenomena
was considerably promoted.
For instance, the roles of crystalline electric field (CEF) potentials
were vigorously discussed.
\cite{Koga1,Koga2,Kusunose1,Kusunose2,OSakai,Koga3,Koga4,Miyake1,Miyake2,Koga5,Miyake3,Miyake4}
In order to observe the two-channel Kondo effect,
first, experiments were performed in cubic U compounds
and then Pr compounds with a $\Gamma_3$ non-Kramers doublet
ground state were extensively investigated.
Recently, in PrT$_2$X$_{20}$ compounds,\cite{Review}
there were significant advances to grasp signs of non-Fermi liquid behavior.
\cite{Sakai,Onimaru1,Onimaru2,Higashinaka}
Theoretical research on this issue was also performed.\cite{Tsuruta,Kusunose3}

At present, research on the two-channel Kondo phenomena
is nearly equivalent to that on the quadrupole two-channel Kondo effect.
However, the magnetic two-channel Kondo effect should also be discussed in actual materials,
when we come back to the original idea of Nozi\'eres and Blandin.
In addition, when we try to observe the two-channel Kondo effect,
the number of candidate materials is limited; thus, it is necessary to develop
Pr compounds with a $\Gamma_3$ non-Kramers doublet ground state.
We believe that it is meaningful to push forward the research frontier
of the two-channel Kondo physics to other rare-earth compounds.

In this study, we suggest that Nd compounds can provide
a new stage of two-channel Kondo phenomena.
We numerically analyze a seven-orbital impurity Anderson model
hybridized with $\Gamma_8$ conduction electrons
for the case with three local $f$ electrons corresponding to a Nd$^{3+}$ ion.
Then, we find a residual entropy of $0.5 \log 2$
as a clear signal of the two-channel Kondo effect,
for the case of the local $\Gamma_6$ ground state.
By analyzing the $\Gamma_6$ state on the basis of the $j$-$j$ coupling scheme,
we propose an extended $s$-$d$ model to explain the present result.
Finally, we provide a few comments on the candidate materials
to detect the two-channel Kondo effect.

First, we define the local $f$-electron Hamiltonian as
\begin{equation}
\label{Hloc}
\begin{split}
  H_{\rm loc} &=\sum_{m_1 \sim m_4}\sum_{\sigma,\sigma'}
  I_{m_1m_2,m_3m_4}
  f_{m_1\sigma}^{\dag}f_{m_2\sigma'}^{\dag}
  f_{m_3\sigma'}f_{m_4\sigma}\\
 &+ \lambda \sum_{m,\sigma,m',\sigma'}
   \zeta_{m,\sigma;m',\sigma'} f_{m\sigma}^{\dag}f_{m'\sigma'}\\
 &+ \sum_{m,m',\sigma} B_{m,m'}
      f_{m \sigma}^{\dag} f_{m' \sigma}
+E_f n,
\end{split}
\end{equation}
where $f_{m\sigma}$ is the annihilation operator
for a local $f$ electron with spin $\sigma$ and $z$-component $m$
of angular momentum $\ell=3$,
$\sigma=+1$ ($-1$) for up (down) spin,
$I$ indicates Coulomb interactions,
$\lambda$ is the spin-orbit coupling,
$B_{m,m'}$ denotes the CEF potentials,
$E_f$ is the $f$-electron level,
and $n$ denotes the local $f$-electron number.

The Coulomb interaction $I$ is expressed as
\begin{equation}
I_{m_1m_2,m_3m_4} = \sum_{k=0}^{6} F^k c_k(m_1,m_4)c_k(m_2,m_3),
\end{equation}
where $F^k$ indicates the Slater-Condon parameter and
$c_k$ is the Gaunt coefficient.\cite{Slater}
The sum is limited by the Wigner-Eckart theorem to
$k=0$, $2$, $4$, and $6$.
Although the Slater-Condon parameters of a material should be determined
from experimental results,
here, we set the ratio as
\begin{equation}
  F^0/10=F^2/5=F^4/3=F^6=U,
\end{equation}
where $U$ is the Hund rule interaction among $f$ orbitals.

Each matrix element of $\zeta$ is given by
\begin{equation}
\begin{split}
\zeta_{m,\sigma;m,\sigma}&=m\sigma/2,\\
\zeta_{m+\sigma,-\sigma;m,\sigma}&=\sqrt{\ell(\ell+1)-m(m+\sigma)}/2,
\end{split}
\end{equation}
and zero for other cases.
The CEF potentials for $f$ electrons from ligand ions are given
in the table of Hutchings for the angular momentum $\ell=3$.\cite{Hutchings}
For a cubic structure with $O_{\rm h}$ symmetry,
$B_{m,m'}$ is expressed by two CEF parameters,
$B_4^0$ and $B_6^0$, as
\begin{equation}
\begin{split}
    B_{3,3}&=B_{-3,-3}=180B_4^0+180B_6^0, \\
    B_{2,2}&=B_{-2,-2}=-420B_4^0-1080B_6^0, \\
    B_{1,1}&=B_{-1,-1}=60B_4^0+2700B_6^0, \\
    B_{0,0}&=360B_4^0-3600B_6^0, \\
    B_{3,-1}&=B_{-3,1}=60\sqrt{15}(B_4^0-21B_6^0),\\
    B_{2,-2}&=300B_4^0+7560B_6^0.
\end{split}
\end{equation}
Note the relation $B_{m,m'}=B_{m',m}$.
Following the traditional notation,\cite{LLW}
we define $B_4^0$ and $B_6^0$ as
\begin{equation}
    B_4^0=Wx/F(4),~B_6^0=W(1-|x|)/F(6),
\end{equation}
where $x$ specifies the CEF scheme for the $O_{\rm h}$ point group,
while $W$ determines the energy scale for the CEF potential.
We choose $F(4)=15$ and $F(6)=180$ for $\ell=3$.\cite{Hutchings}

Now, we consider the case of $n=3$ by appropriately adjusting the value of $E_f$.
As $U$ denotes the magnitude of the Hund rule interaction among $f$ orbitals,
it is reasonable to set $U=1$ eV.
The magnitude of $\lambda$ varies between 0.077 and 0.36 eV
depending on the type of lanthanide ions.
For a Nd$^{3+}$ ion, $\lambda$ is $870-885$ cm$^{-1}$.\cite{Carnall}
Thus, we set $\lambda=0.11$ eV.
Finally, the magnitude of $W$ is typically
of the order of millielectronvolts, although it depends on the material.
Here, we simply set $|W|=10^{-3}$ eV.

\begin{figure}[t]
\centering
\includegraphics[width=8.0truecm]{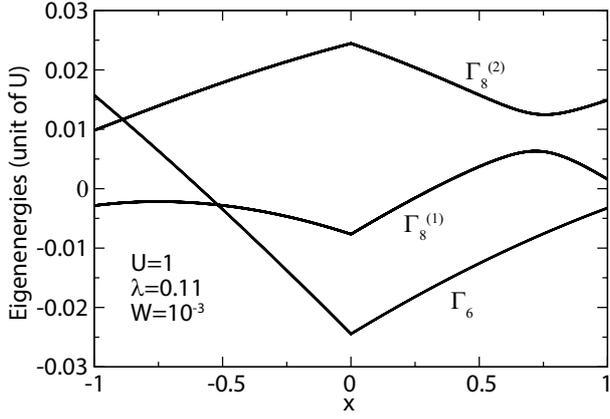}
\caption{Low-lying eigenenergies of $H_{\rm loc}$ versus $x$ for $n=3$.}
\end{figure}

In Fig.~1, we depict curves of ten low-lying eigenenergies of $H_{\rm loc}$
for $n=3$ since the ground-state multiplet for $W=0$ is characterized by $J=9/2$,
where $J$ denotes the total angular momentum of multi-$f$-electron state.
We appropriately shift the origin of the energy
to show all the curves in the present energy range.
We emphasize that the results are almost the same as those of
the $LS$ coupling scheme.\cite{LLW}
For the case of $W>0$, we find the $\Gamma_8^{(1)}$ ground state for $x \le -0.5$,
while the $\Gamma_6$ ground state is observed for $x \ge -0.5$.
However, for $W<0$, the $\Gamma_6$ ground state appears only in
the vicinity of $x=-1.0$.
For the wide range of $-0.9<x \le 1$, we obtain another $\Gamma_8^{(2)}$ ground state.

Now, we include $\Gamma_8$ conduction bands hybridized with localized $f$ electrons.
For the purpose, it is convenient to transform the $f$-electron basis in $H_{\rm loc}$
from $(m, \sigma)$ to $(j, \mu, \tau)$,
where $j$ denotes the total angular momentum of one $f$-electron state,
$\mu$ indicates the irreducible representation of $O_{\rm h}$ point group,
and $\tau$ denotes the pseudo-spin to distinguish the Kramers degenerate state.
For $j=7/2$ octet, we have two doublets ($\Gamma_6$ and $\Gamma_7$)
and one quartet ($\Gamma_8$), while for $j=5/2$ sextet,
we obtain one doublet ($\Gamma_7$) and one quartet ($\Gamma_8$).
In the present case, we consider the hybridization between
$\Gamma_8$ conduction electrons and the $\Gamma_8$ quartet of $j=5/2$.

Then, the seven-orbital Anderson model is expressed as
\begin{equation}
  H \!=\! \! \sum_{\mib{k},\mu,\tau} \! \varepsilon_{\mib{k}}
    c_{\mib{k}\mu\tau}^{\dag} c_{\mib{k}\mu\tau}
   \! +\! \! \sum_{\mib{k},\mu,\tau} \! V(c_{\mib{k}\mu\tau}^{\dag}\tilde{f}_{5/2\mu\tau}+{\rm h.c.})
   \! +\! \tilde{H}_{\rm loc},
\end{equation}
where $\varepsilon_{\mib{k}}$ is the dispersion of a conduction electron with wave vector $\mib{k}$,
$c_{\mib{k}\mu\tau}$ is the annihilation operator of a $\Gamma_8$ conduction electron,
$\mu$ (=$\alpha$ and $\beta$) distinguishes the $\Gamma_8$ quartet,
$\tau$ (=$\uparrow$ and $\downarrow$) is the pseudo-spin,
${\tilde f}_{j\mu\tau}$ is the annihilation operator of a localized $f$ electron
expressed by the bases of $(j, \mu, \tau)$,
$V$ is the hybridization between conduction and localized electrons,
and $\tilde{H}_{\rm loc}$ is obtained from $H_{\rm loc}$ by
the transformation of the $f$-electron basis from $(m, \sigma)$ to $(j, \mu, \tau)$.

\begin{figure}[t]
\centering
\includegraphics[width=8.0truecm]{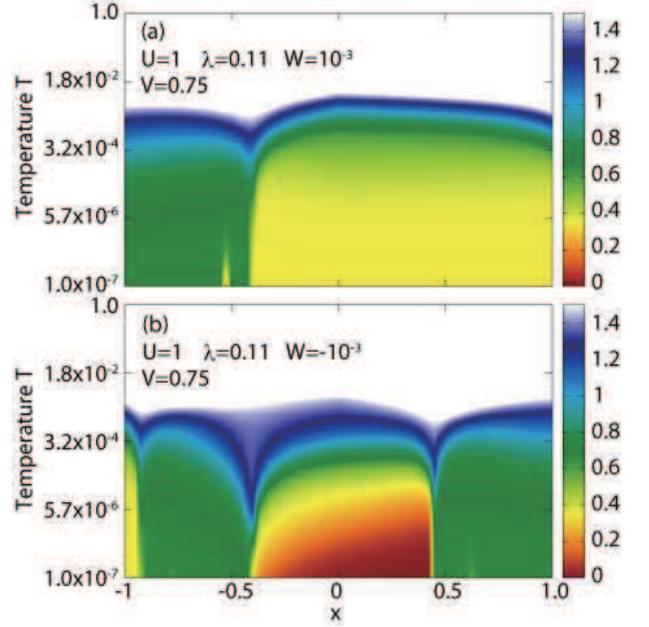}
\caption{(Color online) Contour color map of the entropy on the plane of $x$ and $T$
for (a) $W=10^{-3}$ and (b) $W=-10^{-3}$.
Note that $T$ is given in a logarithmic scale.}
\end{figure}

In this study, we analyze the model by employing
a numerical renormalization group (NRG) method.\cite{NRG1,NRG2}
We introduce a cut-off $\Lambda$ for the logarithmic discretization of
the conduction band.
Owing to the limitation of computer resources,
we keep $M$ low-energy states.
Here, we use $\Lambda=5$ and $M=4,000$.
In the following calculations, the energy unit is $D$, which is
a half of the conduction band width.
Namely, we set $D=U=1$ eV in this calculation.
In the NRG calculation,
the temperature $T$ is defined as $T=\Lambda^{-(N-1)/2}$
in the present energy unit,
where $N$ is the number of renormalization steps.

In Fig.~2(a), we show the contour color map of the entropy for $W=10^{-3}$ and $V=0.75$.
To visualize precisely the behavior of entropy, we define the color of the entropy
between $0$ and $1.5$, as shown in the right color bar.
We immediately notice that an entropy of $\log 2$ (green region) appears at low temperatures
for $-1.0<x < -0.4$, while an entropy of $0.5\log 2$ (yellow region) is found for $-0.4<x<1.0$.
The region with an entropy of $0.5\log 2$ almost corresponds to that of
the $\Gamma_6$ ground state in comparison with Fig.~1,
although we find a small difference between them around $x \sim 0.5$.
The residual entropies, $0.5\log 2$ and $\log 2$, are eventually released
at extremely low temperatures in the numerical calculations.
Approximately at $x=-0.5$, the release of an entropy of $\log 2$ seems to occur
at relatively high temperatures.
This is considered to be related with the accidental degeneracy of
$\Gamma_6$ and $\Gamma_8^{(1)}$ states.
In any case, the details on the entropy behavior at low temperatures
will be discussed elsewhere in the future.

In Fig.~2(b), we show the contour color map of the entropy for $W=-10^{-3}$ and $V=0.75$.
Again, we find a residual entropy of $0.5\log 2$ in the vicinity of $x=-1.0$,
just corresponding to the region of the $\Gamma_6$ ground state for $W<0$,
as observed in Fig.~1.
For $-0.9<x<1$, the CEF ground state is $\Gamma_8^{(2)}$,
but depending on the excited states, the entropy behavior is different.
We find the singlet ground state between $-0.5 < x <0.5$,
while a residual entropy of $\log 2$ appears for $-0.9 < x < -0.5$ and $0.5<x<1$.
For the case of the $\Gamma_8^{(2)}$ ground state,
we do not find a residual entropy of $0.5 \log 2$ even at a certain point of $x$.
Thus, from Figs.~2(a) and 2(b),
we conclude that a residual entropy of $0.5\log 2$
appears for the case of the $\Gamma_6$ ground state for $f^3$ systems.
The Kondo effect for the case of the $\Gamma_8^{(2)}$ ground state
is considered to be related to that in the model with an impurity spin
hybridized with conduction electrons with spin $3/2$.\cite{Cox3,Hattori}
The point will be also discussed elsewhere in the future.

\begin{figure}[t]
\centering
\includegraphics[width=8.0truecm]{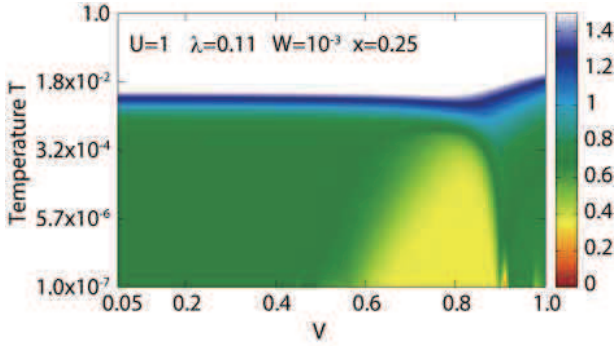}
\caption{(Color online) Contour color map of the entropy on the $(V, T)$ plane
for $x=0.25$ and $W=10^{-3}$.}
\end{figure}

Next we discuss the $V$ dependence of the entropy.
In Fig.~3, we show the contour color map of the entropy on the $(V,T)$ plane
for $x=0.25$ and $W=10^{-3}$ with the $\Gamma_6$ local ground state.
We emphasize that the $0.5\log 2$ entropy does not appear
only at a certain value of $V$,
but it can be observed in the wide region of $V$
such as $0.6 < V < 0.9$ in the present temperature range.
This behavior is different from that in the non-Fermi liquid state
because of the competition between the CEF and Kondo-Yosida singlets
for $f^2$ systems.\cite{Miyake1,Miyake3,Miyake4}
Additionally, the two-channel Kondo effect appears for relatively
large values of $V$ in the energy scale of $U=D=1$ eV.

We believe that the two-channel Kondo effect is confirmed to occur
for the case of $n=3$ with the local $\Gamma_6$ ground state
in the NRG calculation for the seven-orbital Anderson model.
However, it is difficult to describe the electronic state from a microscopic viewpoint
as all $f$ orbitals are included in the present calculations.
Thus, it is desirable to consider the effective model including only $j=5/2$ states
to grasp the essential point of the electronic states.\cite{Hotta1}
For the purpose, we exploit the $j$-$j$ coupling scheme to derive the effective potentials
and interactions among $j=5/2$ states by the perturbation expansion in terms of
$1/\lambda$.\cite{Hotta2}
Then, we perform the NRG calculations for the three-orbital Anderson model
hybridized with $\Gamma_8$ conduction bands
by using the same parameters as those in Fig.~2(a).
Then, we obtain almost the same contour map of entropy
on the $(x,T)$ plane (not shown here)
by using the $j$-$j$ coupling scheme with the effective interactions.

\begin{figure}[t]
\centering
\includegraphics[width=8.0truecm]{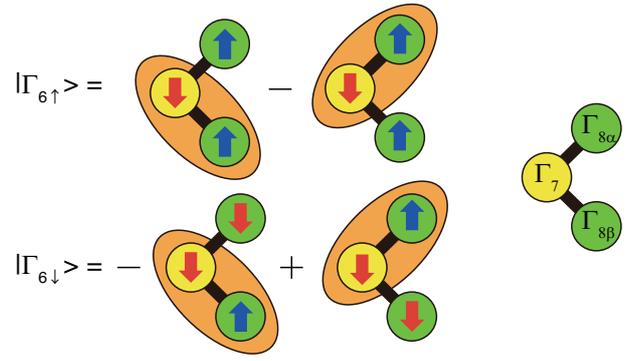}
\caption{(Color online) 
Schematic views of $\Gamma_6$ states composed of the three pseudo-spins
on $\Gamma_7$, $\Gamma_{8\alpha}$, and $\Gamma_{8\beta}$ orbitals of $j=5/2$.
The oval denotes the singlet between $\Gamma_7$ and $\Gamma_8$ states.
The right figure shows the configuration of $\Gamma_7$, $\Gamma_{8\alpha}$,
and $\Gamma_{8\beta}$ orbitals.
}
\end{figure}

Now we consider the local $\Gamma_6$ state on the basis of the $j$-$j$ coupling scheme.
After some algebraic calculations, we express $\Gamma_6$ states
by using the three spins on $\Gamma_7$ and $\Gamma_8$ orbitals,
as schematically shown in Fig.~4.
Namely, we obtain~\cite{Kubo}
\begin{equation}
\begin{split}
|\Gamma_6, \uparrow \rangle &= \frac{1}{\sqrt{3}}
\left( {\tilde f}^{\dag}_{5/2 \alpha \uparrow} |S_{78\beta}\rangle
-{\tilde f}^{\dag}_{5/2 \beta \uparrow} |S_{78\alpha}\rangle \right),\\
|\Gamma_6, \downarrow \rangle &= \frac{1}{\sqrt{3}}
\left( -{\tilde f}^{\dag}_{5/2 \alpha \downarrow} |S_{78\beta}\rangle
+{\tilde f}^{\dag}_{5/2 \beta \downarrow} |S_{78\alpha}\rangle \right),
\end{split}
\end{equation}
where $|S_{78\mu}\rangle$ denotes the singlet state, given by
\begin{equation}
|S_{78\mu}\rangle = \frac{1}{\sqrt{2}}
\left( {\tilde f}^{\dag}_{5/2 \gamma \uparrow}{\tilde f}^{\dag}_{5/2 \mu \downarrow}
-{\tilde f}^{\dag}_{5/2 \gamma \downarrow}{\tilde f}^{\dag}_{5/2 \mu \uparrow}
\right) |0\rangle.
\end{equation}
Here, ${\tilde f}_{5/2\gamma\tau}$ is the annihilation operator of $\Gamma_7$ electron
and $|0\rangle$ denotes the vacuum.
The main component of the $\Gamma_3$ non-Kramers doublet state of $n=2$
is expressed by $|S_{78\alpha}\rangle$ and $|S_{78\beta}\rangle$.
Thus, the $\Gamma_6$ states of $n=3$ are obtained
by the addition of one $\Gamma_8$ electron to $\Gamma_3$ states of $n=2$.
We intuitively understand that the pseudo-spin properties of $\Gamma_6$
originate from those of $\Gamma_8$ electrons.

On the basis of local $\Gamma_6$ states composed of the three pseudo-spins,
we obtain an extended $s$-$d$ model, given by
\begin{equation}
  H \!=\! \! \! \sum_{\mib{k},\mu,\tau} \! \! \varepsilon_{\mib{k}}
    c_{\mib{k}\mu\tau}^{\dag} c_{\mib{k}\mu\tau}
   \! +\! J_1 \! \! \! \sum_{\mu=\alpha,\beta} \! \! \mib{s}_{\mu} \cdot \mib{S}_{8\mu}
   \! +\! J_2 \! \! \! \sum_{\mu=\alpha,\beta} \! \! \mib{S}_{7} \cdot \mib{S}_{8\mu},
\end{equation}
where $J_1$ is the Kondo exchange coupling,
$J_2$ denotes the effective negative Hund rule coupling
among $\Gamma_7$ and $\Gamma_8$ orbitals,\cite{Kubo}
$\mib{s}_{\mu}$ denotes the conduction electron spin for the $\mu$ orbital,
and $\mib{S}_{7}$ and $\mib{S}_{8\mu}$ indicate the local spin
on $\Gamma_7$ and $\Gamma_{8\mu}$ orbitals, respectively.

In Fig.~5, we show the typical results of the entropy and specific heat
of the extended $s$-$d$ model with $J_1=0.5$ and $J_2=0.1$.
We clearly find a residual entropy of $0.5 \log 2$ at low temperatures,
suggesting the emergence of the two-channel Kondo effect.
In contrast to the well-known two-channel Kondo model,
we observe a remnant of the plateau of $\log 2$
before entering the two-channel Kondo region.
Corresponding to the change of the entropy,
we observe a small peak in the specific heat.
Here, we show only the results for $J_1=0.5$ and $J_2=0.1$,
but the two-channel Kondo behavior can be widely observed
for positive values of $J_1$ and $J_2$.

\begin{figure}[t]
\centering
\includegraphics[width=8.0truecm]{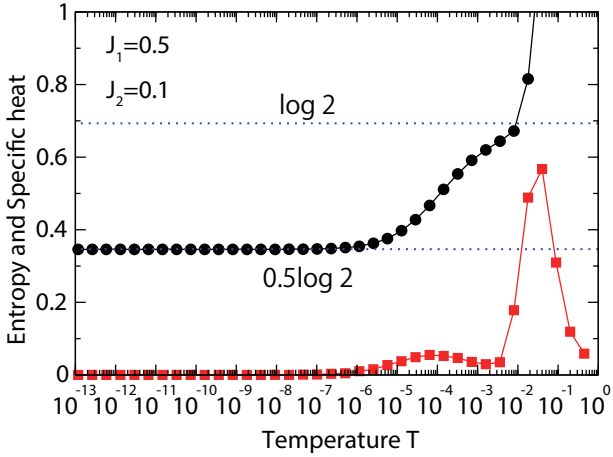}
\caption{(Color online) Entropy and specific heat of the extended $s$-$d$ model for $J_1=0.5$ and $J_2=0.1$.}
\end{figure}

Now, we briefly discuss the candidate $f^3$ materials to observe the
two-channel Kondo behavior.
A simple way is to search Nd cubic compounds with $\Gamma_6$ ground states.
For the purpose, it is convenient to synthesize Nd compounds with the same crystal structure
as that of Pr compounds with a $\Gamma_3$ non-Kramers ground state
from the discussion on the CEF states for $n=2$ and $3$.
As we have remarked in Fig.~3, to observe the two-channel Kondo effect in $f^3$ systems,
it is necessary to consider relatively large hybridization,
suggesting that a rare-earth ion should be surrounded by many ligand ions.
In this sense, good candidates are considered to be Nd 1-2-20 compounds,
such as NdIr$_2$Zn$_{20}$,\cite{Nd1}
NdRh$_2$Zn$_{20}$, NdV$_2$Al$_{20}$,\cite{Nd2}
and NdTi$_2$Al$_{20}$,\cite{Nd3,Nd4}
since corresponding Pr compounds are known to exhibit
$\Gamma_3$ non-Kramers doublets.\cite{Review}

Among them, the $\Gamma_6$ ground state has been
confirmed for NdIr$_2$Zn$_{20}$,\cite{Nd1}
but the signals of the two-channel Kondo effect have not been reported.
At low temperatures, antiferromagnetic phases
have been observed for NdIr$_2$Zn$_{20}$~\cite{Nd1}
and NdTi$_2$Al$_{20}$,\cite{Nd4}
while a ferromagnetic phase has been found for NdV$_2$Al$_{20}$.\cite{Nd2}
Thus, a high pressure can be applied to such magnetic phases
since we expect a chance to observe the two-channel Kondo behavior
if we obtain a metallic phase through the quantum critical point
under a high pressure.

Another candidate may be found in Np cubic compounds
with a Np$^{4+}$ ion including three $5f$ electrons
because in general, the itinerant nature of $5f$ electrons is large
in comparison with that of $4f$ electrons.
However, as the treatment of Np compounds is strictly limited,
it may be difficult to find the two-channel Kondo behavior
in Np cubic compounds.

In summary, we found the two-channel Kondo effect
in the seven-orbital impurity Anderson model
hybridized with $\Gamma_8$ conduction electrons
for the case of $n=3$ with the local $\Gamma_6$ ground state.
To detect the two-channel Kondo effect emerging from Nd ions,
we proposed to perform the experiments using Nd 1-2-20 compounds.

The author thanks Y. Aoki, K. Hattori, R. Higashinaka, K. Kubo, and T. Matsuda
for discussions on heavy-electron systems.
This work was supported by JSPS KAKENHI Grant Number JP16H04017. 
The computation in this work was done using the facilities of the
Supercomputer Center of Institute for Solid State Physics, University of Tokyo.


\end{document}